\documentclass[intlimits,twoside,a4paper]{article}
\usepackage{amsmath,amssymb}
\usepackage{graphicx}
\usepackage{wrapfig}
\usepackage[T2A]{fontenc}
\usepackage[cp1251]{inputenc}

\usepackage{cmpj}

\newcommand{\ve}{\varepsilon}

\issue{2011}{14}{2}{23602}

\doinumber{10.5488/CMP.14.23602}

\title[Morphology change of the silicon surface]%
{Morphology change of the silicon surface induced by Ar$^+$ ion beam sputtering}%

\author[V.O. Kharchenko, D.O. Kharchenko]{V.O. Kharchenko, D.O. Kharchenko}
\address{Institute of Applied Physics, National
Academy of Sciences of Ukraine, \\58~Petropavlivska~Str.,
40030~Sumy, Ukraine}

\date{Received April 19, 2011, in final form June 6, 2011}

\authorcopyright{V.O. Kharchenko, D.O. Kharchenko, 2011}

\begin{document}

\maketitle

\begin{abstract}
Two-level modeling for nanoscale pattern formation on silicon
target by Ar$^+$ ion sputtering is presented. Phase diagram
illustrating possible nanosize surface patterns is discussed.
Scaling characteristics for the structure wavelength dependence
versus incoming ion energy are defined. Growth and roughness
exponents in different domains of the phase diagram are obtained.
\pacs 68.55.-a, 68.35.Ct, 79.20.Rf, 81.16.Rf
\end{abstract}

\section{Introduction}

It is well known that low and medium energy ion sputtering may
induce a fabrication of periodic nanoscale structures on an
irradiated surface~\cite{cite39}. Depending on the sputtered
substrate characteristics and sputtering conditions, different
types of nanoscale structures such as ripples, nanoholes and
nanodots can grow on a target during ion beam
sputtering~\cite{cite10,cite11,cite12,cite21,cite26}. These
patterns have been found on both amorphous and crystalline
materials including insulators, semiconductors and metals (see
reference~\cite{cite38} and citations therein). Main theoretical
models describing ripple formation are based on the results of
famous works by Bradley and Harper~\cite{cite12}, Kardar et
al.~\cite{cite13}, Wolf and Villian~\cite{cite14}, and Kuramoto et
al.~\cite{cite15}. The main control parameters in these models reduced
to surface tensions, tilt-dependent erosion rates and diffusion
constants are determined by sputtered substrate characteristics
and sputtering conditions (see for example~\cite{cite38}).

Among theoretical investigations there are a lot of experimental data
manifesting a large class of patterns formed due to the self-organization
process. It was experimentally shown that the main properties of pattern formation
processes depend on ion-beam parameters such as ion flux, energy of deposition,
angle of incidence, and temperature of the substrate (target). Therefore, to study the ion beam sputtering
processes theoretically one needs to determine the mentioned
parameters of the model according to the physical conditions related to
concrete materials.

One of the most frequently used materials for ion beam
sputtering is silicon because it is the mainstream material in
modern microelectronic industry and it is readily available with high
purity and quality. Nanostructuring of silicon has received much
attention due to its potential application in developing the Si light
sources~\cite{cite40}. Various techniques such as acid etching,
ion implantation, reactive evaporation, chemical vapor deposition
and molecular beam epitaxy have been used in developing the Si
nanomaterials (porous Si, Si nanocrystal-doped dielectrics and Si
quantum dots) (see reference~\cite{cite40} and citation therein).

In this paper we study the properties of the formation of nanoscale patterns
on a silicon target sputtered by Ar$^+$ ions. To this end, we use a
two-level scheme, based on Monte-Carlo simulations and the modified
Bradley-Harper theory. In the first approach we compute the ion energy
dependent penetration depth, widths of the ion energy distribution
and sputtering yield. Next, we exploit these characteristics as
input data for the continuum approach describing the evolution of the
surface height field. We define the domains of values for the angle of
incidence and ion energy where different nanoscale structures can
be formed. The dynamics of nanoscale pattern formation is discussed.
We will show that at fixed values for the incidence angle, one has two
scaling exponents for wavelength related to small and large values
for ion energy. In addition, we obtain roughness and growth
exponents.

The work is organized in the following manner. In section 2 we
present the theoretical model in the framework of the modified
Bradley-Harper approach. In section 3 using Monte-Carlo modeling
we compute the main characteristics incorporated into the continuum
theory. The related phase diagram, the dynamics of the formation of nanoscale structures, the ion energy dependent wavelength and scaling properties
of the surface morphology are discussed in section 4. We conclude
in the last section.

\section{Theoretical model}

Let us consider a $d$-dimensional substrate and denote with
$\mathbf{r}$ the $d$-dimensional vector locating a point on it.
The surface is described at each time $t$ by the height
$z=h(\mathbf{r},t)$. If we assume that the surface morphology
changes during ion sputtering, then we can use the model for the
surface growth proposed by Bradley and Harper~\cite{cite12} and
further developed by Cuerno and Barabasi~\cite{cite10}. We
consider the system where the direction of the ion beam lies in
$x-z$ plane at an angle $\theta$ to the normal of the uneroded
surface. Following the standard approach one assumes that an
averaged energy deposited at the surface (let say at point $O$), due
to the ion arriving at the point $P$ in the solid, follows the
Gaussian distribution~\cite{cite12}
$E(\mathbf{r})=\left[\varepsilon/(2\pi)^{3/2}\sigma\mu^2\right]
\exp\left[-z^2/2\sigma^2-(x^2+y^2)/2\mu^2\right]$; $\varepsilon$
denotes the kinetic energy of the arriving ion, $\sigma$ and $\mu$
are the widths of distribution in directions parallel and
perpendicular to the incoming beam. Parameters $\sigma$ and $\mu$
depend on the target material and can vary with physical
properties of the target and incident energy. The erosion velocity
at the surface point $O$ is described by the formula
$v=p\int_\mathcal{R}{\rm
d}\mathbf{r}\Phi(\mathbf{r})E(\mathbf{r})$, where integration is
provided over the range of the energy distribution of all ions;
here $\Phi(\mathbf{r})$ are the corrections for the local slope
dependence of the uniform flux $J$. The material constant $p$ is
defined as: $p=3/(4\pi^2)(NU_0C_0)^{-1}$, where $U_0$ and $C_0$
are the surface binding energy and the constant proportional to the square
of the effective radius of the interatomic interaction potential,
respectively~\cite{cite18}. The general expression for the local
flux for surfaces with non-zero local curvature is~\cite{cite17}:
\[\Phi(x,y,h)=J\cos\left[\arctan\left(\sqrt{(\nabla_x
h)^2+(\nabla_y h)^2}\right)\right].
\]
 Hence, the dynamics of the surface height is defined by the relation
$\partial_t h\simeq-v\left(\theta-\nabla_x h, \nabla_x^2 h,
\nabla_y^2 h\right)$ and is given by the equation $\partial_t
h\simeq-v(\theta)\sqrt{1+(\nabla h)^2}$, where
$0<\theta<\pi/2$~\cite{cite12,CT94,cite10,cite13,cite11}. The
linear term expansion yields $\partial_t
h=-v_0+\gamma\nabla_xh+\nu_\alpha\nabla_{\alpha\alpha}^2h$; where
$\nabla=\partial/\partial \mathbf{r}$,
$\nabla_\alpha=\partial/\partial \alpha$, $\alpha=\{x,y\}$. Here
$v_0$ is the surface erosion velocity; $\gamma=\gamma(\theta)$ is
a constant that describes the slope depending erosion;
$\nu_\alpha=\nu_\alpha(\theta)$ is the effective surface tension
generated by erosion process in $\alpha$ direction.

If one assumes that the surface current is driven by differences
in chemical potential $\mu$, then the evolution equation for the
field $h$ should take into account the term $-\nabla\cdot
\mathbf{j}_s$ in the right hand side, where $\mathbf{j}_s=
K\nabla(\nabla^2h)$ is the surface current; $K>0$ is the
temperature dependent surface diffusion constant. If the surface
diffusion is thermally activated, then we have
$K=D_s\kappa\rho/n^2T$, where $D_s=D_0\exp\left(-E_a/T\right)$ is
the surface self-diffusivity ($E_a$ is the activation energy for
surface diffusion), $\kappa$ is the surface free energy, $\rho$ is
the areal density of diffusing atoms, $n$ is the number of atoms
per unit volume in the amorphous solid. This term in the dynamical
equation for $h$ is relevant in high temperature limit which will
be studied below.

Assuming that the surface varies smoothly, we neglect spatial
derivatives of the height $h$ of third and higher orders in the
slope expansion. Taking into account nonlinear terms in the slope
expansion of the surface height dynamics, we arrive at the
equation for the quantity $h'=h+v_0t$ of the
form~\cite{cite12,cite10}
\begin{equation}\label{Aeq1}
\partial_t h=\gamma\nabla_xh+\nu_\alpha\nabla_{\alpha\alpha}^2h+\frac{1}{2}\lambda_\alpha(\nabla_\alpha h)^2-
K\nabla^2(\nabla^2h)+
 \xi(x,y,t),\qquad \alpha=\{x,y\},
\end{equation}
where we drop the prime for convenience. Here we introduce the
uncorrelated white Gaussian noise $\xi$ with zero mean mimicking
the randomness resulting from the stochastic nature of the ion arrival
to the surface. In equation~(\ref{Aeq1}) the effective surface
tensions $\nu_x$ and $\nu_y$ generated by the IBS, the tilt-dependent
erosion rates $\lambda_x$ and $\lambda_y$ are defined through the
incident angle $\theta$, penetration depth of incident ion $a$,
distribution widths $\sigma$, $\mu$ and the sputtering yield $Y_0$ as
follows \cite{cite12,cite10}:
\begin{eqnarray} \label{Aeq3}
 \gamma&=&F_0\frac{s}{f^2}\left[a_\sigma^2a_\mu^2c^2\left(a_\sigma^2-1\right)
 -a_\sigma^4s^2\right],\nonumber\\
 \nu_x&=&F_0a\frac{a_\sigma^2}{2f^3}\left(2a_\sigma^4s^4
 -a_\sigma^4a_\mu^2s^2c^2+a_\sigma^2a_\mu^2s^2c^2-a_\mu^4c^4\right),\nonumber\\
 \nu_y&=&-F_0a\frac{c^2a_\sigma^2}{2f}\,, \nonumber\\
 \lambda_x&=&F_0\frac{c}{2f^4}\left\{a_\sigma^8a_\mu^2s^4\left(3+2c^2\right)
 +4a_\sigma^6a_\mu^4c^4s^2-a_\sigma^4a_\mu^6c^4 \left(1+2s^2\right)\right. \nonumber\\
 &&{}-\left.f^2\left[2a_\sigma^4s^2-a_\sigma^2a_\mu^2\left(1+2s^2\right)\right]
 -a_\sigma^8a_\mu^4c^2s^2-f^4\right\},\nonumber \\
 \lambda_y&=&F_0\frac{c}{2f^2}\left(a_\sigma^4s^2+a_\sigma^2a_\mu^2c^2
 -a_\sigma^4a_\mu^2c^2-f^2\right).
\end{eqnarray}
Here we have used the following notations:
\begin{equation} \label{Aeq4}
F_0\equiv \frac{J\ve Y_0pa}{\sigma \mu \sqrt{2\pi f} } \exp
\left(\frac{-a_{\sigma }^{2} a_{\mu }^{2} c^{2} }{2f} \right),
\end{equation}
\begin{equation} \label{Aeq5}
a_\sigma\equiv\frac{a}{\sigma}\,,\qquad
a_\mu\equiv\frac{a}{\mu}\,,\qquad s\equiv \sin(\theta),\qquad
c\equiv\cos(\theta),\qquad f\equiv a_{\sigma }^{2} s^{2} +a_{\mu
}^{2} c^{2}.
\end{equation}

Let us perform the stability analysis for a system with additive
fluctuations. To this end, we average the Langevin equation (\ref{Aeq1}) over
noise and obtain
\begin{equation} \label{GrindEQ1}
{\partial_t} \left\langle h\right\rangle =\gamma \nabla_x\left\langle
h\right\rangle +\nu _{x} \nabla^2_{xx} \left\langle h\right\rangle +\nu _{y}
\nabla^2_{yy} \left\langle h\right\rangle +\frac{\lambda _{x} }{2} \left\langle
\left(\nabla_x h\right)^{2} \right\rangle +\frac{\lambda _{y} }{2} \left\langle
\left(\nabla_y h\right)^{2} \right\rangle -K\nabla ^{4} \left\langle
h\right\rangle.
\end{equation}
Considering the stability of the smooth surface characterized by $\langle h
\rangle=0$, we can rewrite the linearized evolution equation in the standard
form:
\begin{equation} \label{GrindEQ2}
{\partial_t} \left\langle h\right\rangle =\left(\hat{\nu }_{\rm
ef} +\hat{K}_{\rm ef} \right)\left\langle h\right\rangle ,
\end{equation}
with notations
\begin{eqnarray} \label{GrindEQ3}
 \hat{\nu }_{\rm ef} =\gamma {\nabla_x} +\nu _{x}\nabla_{xx}^2
+\nu _{y} \nabla_{yy}^2 \, , \qquad {\hat{K}_{\rm ef} =-K\nabla
^{4} } .
\end{eqnarray}

It is easy to see that equation~(\ref{GrindEQ3}) admits a solution
of the form $\langle h\rangle=A\exp[\ri(k_xx+k_yy-\omega t)+\chi
t]$. Indeed, substituting it into equation~(\ref{GrindEQ3}) and
separating real and imaginary parts we get
\begin{eqnarray} \label{GrindEQ4}\omega &=&-\gamma \left(\theta
\right)k_{x}\,,
\\ \chi&=&-\nu _{x}
\left(\theta \right)k_{x}^{2} -\nu _{y} \left(\theta
\right)k_{y}^{2} -K\left(k_{x}^{2} +k_{y}^{2} \right)^{2}.
\end{eqnarray}
As far as $F_0$\,, $f$, $a$ are positive values, hence one has
$\nu_y<0$, whereas $\nu_x$ can change its sign. Therefore, the
Bradley-Harper model does not provide for stable smooth surface.
Hence, we can conclude that if $\nu_x>0$, then ripples (wave
patterns) appear in $x$-direction. On the contrary, when $\nu_x<0$,
equiaxed structures (nanodots/nanoholes) can be formed on an
eroded surface. In addition, the sign of the product
$\lambda_x\cdot\lambda_y$ can play a crucial role in ripple
formation processes~\cite{cite27}.

For the noiseless nonlinear model~(\ref{Aeq1}) it was shown that
as the sets $\nu_\alpha$ and $\lambda_\alpha$ are the functions of
the angle of incidence $\theta\in [0,\pi/2]$ there are three
domains in the phase diagram $(a_\sigma\,,\theta)$ where $\nu_x$
and $\lambda_x$ change their signs, separately~\cite{cite10}. This
results in the formation of ripples in different directions $x$ or $y$
varying $a_\sigma$ or $\theta$.

One needs to note that in the Bradley-Harper approach describing
the processes of ripple formation on amorphous substrates the
penetration depth can be approximated as $a(\ve)\sim\ve$ leading
to the power-law asymptotics for the wavelength of the ripples
$\Lambda$ versus ion energy as follows:
$\Lambda\sim\ve^{-1/2}$~\cite{cite12}. In the next section,
performing calculations for the sputtering of the silicon target by
Ar$^+$ ions we shall show that there are deviations from these
asymptotics due to power-law dependence of $a$, $\sigma$, $\mu$
versus ion energy $\ve$. Moreover, it will be shown that the
sputtering yield depends on both incident angle $\theta$ end ion
energy $\ve$ in a power-law form. To define $a$, $\sigma$, $\mu$
and the sputtering yield, we shall use Monte-Carlo approach.

\section{Monte-Carlo modelling}

To study the evolution of the silicon surface morphology during Ar$^+$
ion beam sputtering one needs to know the energy of ions and target
characteristics such as: penetration depth of the Ar$^+$ ions into
the silicon target $a$; widths of the distribution in parallel and
perpendicular directions of the incoming beam ($\sigma$ and $\mu$)
and sputtering yield $Y_0$\,. Moreover, to simulate the target
morphology evolution one should define temperature $T$, uniform
flux $J$, atomic density of the target $N$, surface binding energy
$U_0$ and the effective radius of interatomic interaction
potential. Using data from reference~\cite{cite21} at $T=550$~C
one has $K=C'8.49\times10^{3}$~nm$^4$/s, where addimer
concentration is $C'=0.04$ atoms/site ($\approx4\%$ coverage), or,
$C'=0.07$~atoms/nm$^2$. To define the material constant $p$ we shall
use $N\simeq50$~atoms/nm$^3$, $U_0=4.73$~eV. For the effective
radius of interatomic interaction potential, we put the length of
the main diagonal of silicon primitive cell with lattice parameter
$0.5437$~nm. To compute the time evolution of the silicon surface
morphology, we should calculate the effective surface tensions $\nu_x$ and $\nu_y$
generated by the IBS, and the rates $\lambda_x$ and $\lambda_y$ [see
equation~(\ref{Aeq3})]. Following
relations~(\ref{Aeq4}) and~(\ref{Aeq5}) dependent on ion energy,
parameters $a$, $\sigma$ and $\mu$, as far as $Y_0=Y_0(\theta,\ve)$,
can be computed. From experimental point of view, the control
parameters at IBS are the energy of ion beam, off-normal incidence
angle and ion flux. In all our calculations we put
$J=20$~ions/(nm$^2$\,s). Thereafter we vary the ion energy in the
interval $100$~eV$\div10$~keV and use intermediate off-normal
incidence angles $\theta\in [40^{\circ},65^{\circ}]$.

In further study, we use the well-known program codes (TRIM and
SRIM) to calculate the stopping range of ions in matter and
transport range of ions in matter. Description of algorithms and
the basic principles for Monte-Carlo calculation of both the
transport range of ions in matter and the stopping range of ions
in matter can be found in~\cite{book_TRIM}; TRIM and SRIM codes
can be found on the web-site \href{www.srim.org}{www.srim.org}.

\begin{figure}[ht]
\centering\includegraphics[width=8cm]{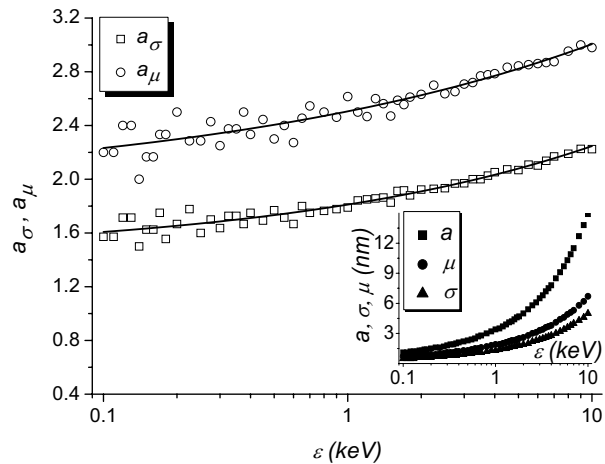}
\caption{Relative penetration depths of the Ar$^+$ ions into the
silicon target versus ion energy (penetration depth and
distribution widths versus ion energy are shown in insertion).}
\label{fig1}
\end{figure}

Values for parameters $a$, $\sigma$ and $\mu$ for silicon target
sputtered by Ar$^+$ ions were obtained with the help of SRIM code
(a program for calculating the stopping range of ions in matter).
Results for relative penetration depths $a_\sigma\equiv a/\sigma$
and $a_\mu\equiv a/\mu$ versus ion energy are shown in
figure~\ref{fig1} (dependencies $a(\varepsilon)$,
$\sigma(\varepsilon)$ and $\mu(\varepsilon)$ are shown in the
insert).
In figure~\ref{fig1} it is seen that longitudinal and transverse
widths $\sigma$ and $\mu$, respectively, as far as $a_\sigma$ and
$a_\mu$ satisfy the following relations, are as follows: $\sigma<\mu$ and
$a_\sigma<a_\mu$\,. In reference~\cite{cite29} it was shown that the
rotated ripple structures formed when $\lambda_x\lambda_y<0$ with
rotation angle $\varphi=\tan^{-1}\sqrt{-\lambda_x/\lambda_y}$ can
be observed at small incidence angles $\theta$ when
$a_\mu<a_\sigma$ ($a_\sigma=1$) and at intermediate and large
$\theta$ when $a_\mu>a_\sigma$\,. Hence, one can expect the
appearance of rotated ripple structures in our system.

It is principally important that dependencies of penetration depth
$a$ and longitudinal and transverse widths $\sigma$ and $\mu$,
respectively, versus ion energy $\ve$ deviate from the linear law
predicted by Bradley and Harper~\cite{cite12}. For a silicon
target sputtered by Ar$^+$ ions we have obtained a power-law
approximation of the form: $\phi(\ve)=A+B\ve^C$, where
$\phi=\{a,\sigma,\mu\}$, constants $A$, $B$ and $C$ are fitting
parameters. So, we can expect that the wavelength dependence
$\Lambda(\ve)\sim\ve^{-\delta}$ can be characterized by the
exponent $\delta\ne1/2$. In our further continuum approach we
shall use the obtained power-law asymptotics for $a$, $a_\sigma$ and
$a_\mu$ from Monte-Carlo simulations.

\begin{figure}[ht]
\includegraphics[width=0.48\textwidth]{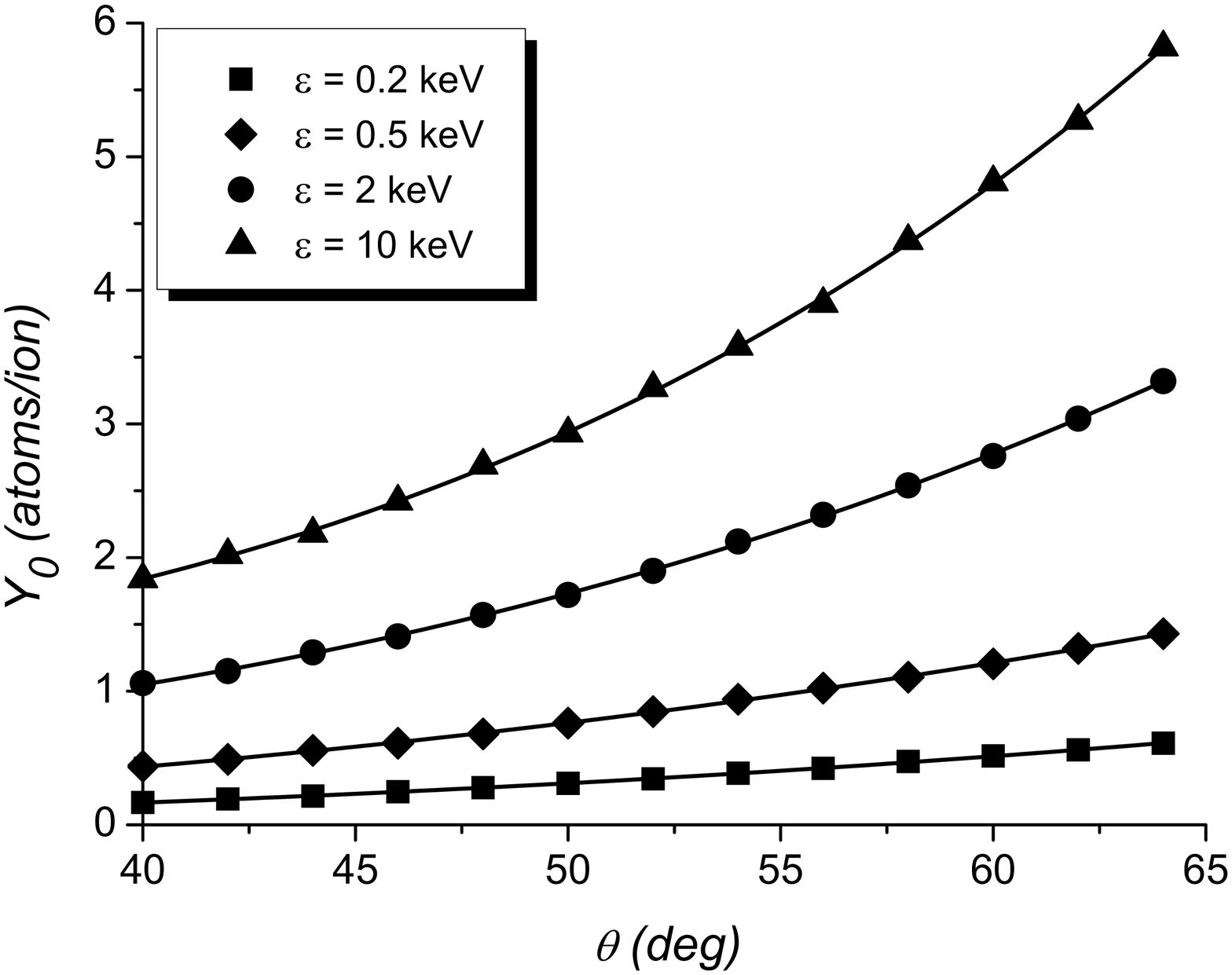}%
\hfill%
\includegraphics[width=0.48\textwidth]{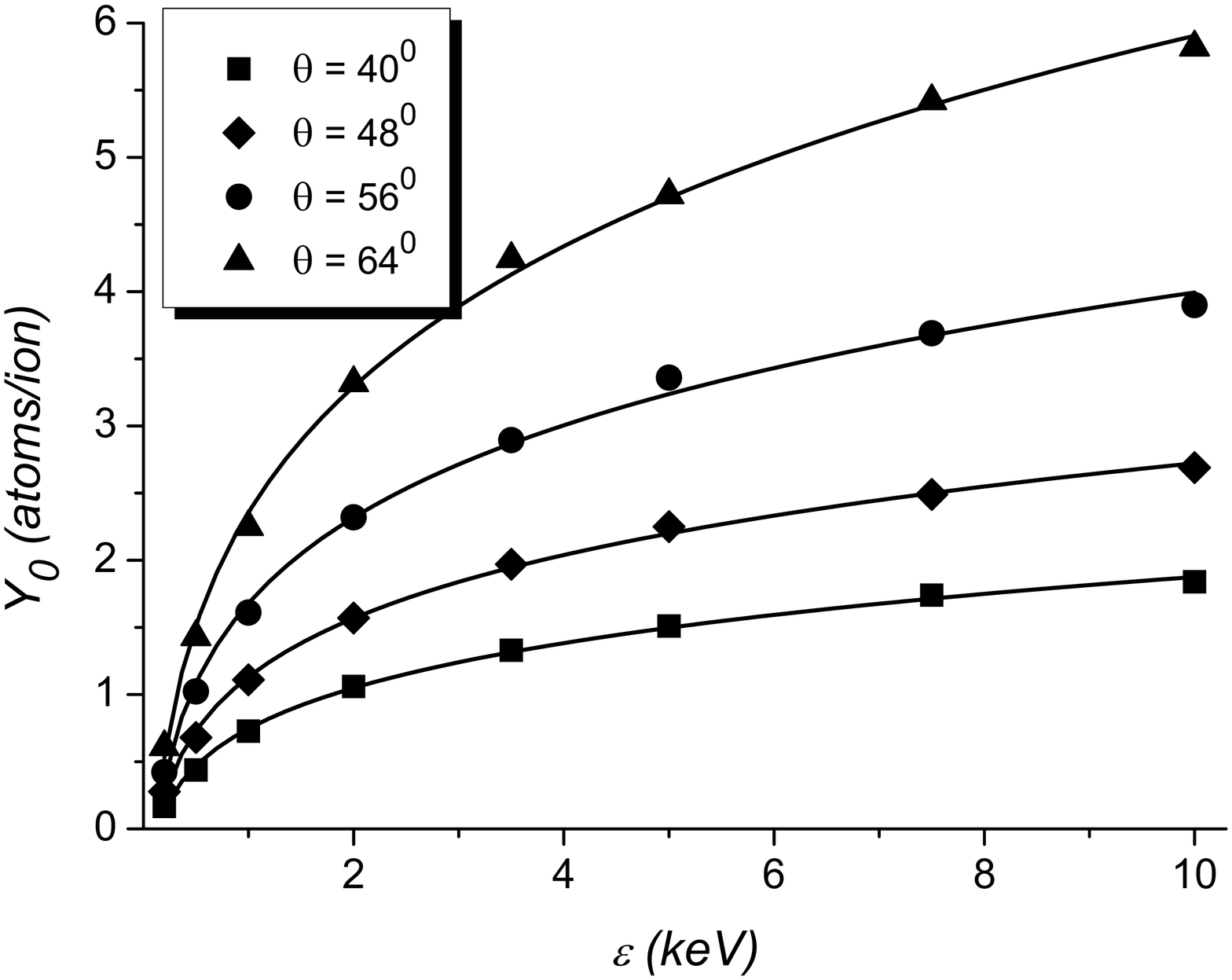}%
\\%
\parbox[t]{0.48\textwidth}{%
\centerline{(a)}%
}%
\hfill%
\parbox[t]{0.48\textwidth}{%
\centerline{(b)}%
}%
\caption{Sputtering yield for Ar$^+$ in Si at (a) fixed ion energy,
(b) fixed incidence angle.} \label{fig2}
\end{figure}

To compute the dependence of the sputtering yield versus ion
energy and angle of incidence we use Monte-Carlo approach realized
in TRIM code (program for the calculation of transport range of ions in
matter). The results of calculations for sputtering yield versus
incident angle at fixed ion energy and sputtering yield versus ion
energy at fixed incident angle are shown in figures~\ref{fig2}~(a)
and \ref{fig2}~(b), respectively.
In figure~\ref{fig2} it is seen that sputtering yield depends on
both ion energy and incidence angle in accordance with a power law
as follows $Y_0(\psi)=A'+B'\psi^{C'}$, where
$\psi=\{\theta,\ve\}$, constants $A'$, $B'$ and $C'$ are fitting
parameters. Therefore, all parameters ($\nu_x$\,, $\nu_y$\,,
$\lambda_x$\,, $\lambda_y$) required to monitor the time evolution of
 silicon surface morphology during IBS are well defined.

\section{Surface morphology change during sputtering}

\subsection{Phase diagram and typical patterns}

\begin{figure}[ht]
\centering
\includegraphics[width=8cm]{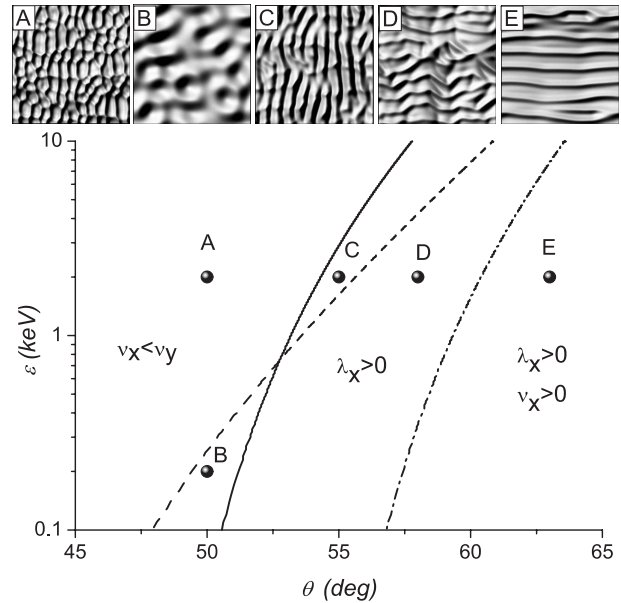}
\caption{Phase diagram and typical surface patterns.} \label{fig3}
\end{figure}

Firstly, let us compute a phase diagram $\ve(\theta)$ defining
domains for different surface patterns of silicon sputtered by
Ar$^+$ ions. To this end, we shall monitor a sign change of surface
tension $\nu_x$ and tilt-dependent erosion rates $\lambda_x$ and
$\lambda_y$ (as it was mentioned above $\nu_y$ is always less than 0). The
corresponding phase diagram indicating possible patterns is shown
in figure~\ref{fig3}. We need to stress that in the related
interval for both the ion energy and the incidence angle except
$\nu_y<0$, one has $\lambda_y<0$.
From figure~\ref{fig3} it follows that plane $(\theta,\ve)$ is
divided by three curves into five domains {\it A}, {\it B}, {\it
C}, {\it D} and {\it E}. If one crosses the dash-dot curve, then
quantity $\nu_x$ changes it sign. Therefore, in the linear regime
at small incidence angles $\theta$ (domain {\it A}), instability of
the silicon surface occurs in both $x$ and $y$ directions due to
$\nu_y<0$ and $\nu_x<0$. In the domain {\it E} (at large $\theta$)
in the linear regime, patterns are stable in $x$-direction due to
$\nu_x>0$. At large times (nonlinear regime) the surface
morphology is governed by nonlinear parameters $\lambda_x$ and
$\lambda_y$\,. Solid curve in figure~\ref{fig3} divides domains
characterized by $\lambda_x<0$ and $\lambda_x>0$. Therefore,
between solid and dash-dot lines only $\lambda_x$ is positive
(domains {\it C} and {\it D}), whereas in the domain {\it E} both
$\nu_x$ and $\lambda_x$ are positive. Dash curve corresponds to
the condition $\nu_x=\nu_y$\,. Hence, before the dash curve
(domains {\it A} and {\it C}) when $\nu_x<\nu_y$, vertical elongated
surface structures should be formed, whereas after the dash curve
(domains {\it B}, {\it D} and {\it E}), the corresponding
structures should be of a horizontal elongated type.

To illustrate typical structures in each domain in
figure~\ref{fig3} we numerically solve equation~(\ref{Aeq1}) on
quadratic lattice $L\times L$ of the linear size $L=256$ with
periodic boundary conditions. Spatial derivatives of the second
and fourth orders were computed according to the standard
finite-difference scheme; the nonlinear term $(\nabla h)^2$ was
computed according to the scheme proposed in
references~\cite{cite33,cite34}. We have used Gaussian initial
conditions taking $\langle h({\bf r},t=0)\rangle=0$ and
$\langle(\delta h)^2\rangle=0.1$; the integration time step is
$\Delta t = 0.005$ and the space step is $\ell=1$.

Typical surface patterns in domains ({\it A}--{\it E}) are shown in
figure~\ref{fig3}. It is seen that on the left hand side of the solid curve when
$\nu_y<0$ and $\nu_x<0$, pattern type of holes is realized (see
snapshots A and B). It follows that patterns realized at high
energy ions are characterized by small size (see snapshot A),
whereas at small $\ve$ one has large-scale patterns (see snapshot
B)\footnote{Dependence of wavelength versus ion energy at fixed
values for incidence angle will be discussed later.}. Moreover,
orientation of holes in points {\it A} and {\it B} is different.
It is defined by a minimal value of both $\nu_x$ and $\nu_y$\,.
Structures, shown by snapshots C and D (ripples) are characterized
by positive value of parameter $\lambda_x$\,, which defines
nonlinear effects in $x$-direction. An orientation of the
corresponding ripples is defined by a minimal value of both
$\nu_x$ and $\nu_y$ as in the previous case. Hence, as far as
$\nu_x<\nu_y$ from the left of the dashed curve in figure~\ref{fig3},
the related patterns in domains {\it A} and {\it C} are elongated
in $y$ direction. On the contrary, in snapshots B, D, and E there are
horizontal elongated structures. Structures in snapshots A, B, C
and D are characterized by instabilities in both $x$ and $y$
directions due to $\nu_x<0$ and $\nu_y<0$. In the domain,
indicated by point {\it E} due to $\nu_x>0$, structures are stable
in $x$ direction.

The obtained phase diagram is in good correspondence with the results of
experimental studies of the dynamics of the surface Si(001) sputtered by
Ar$^+$ ions~\cite{cite22}, where according to the experimentally
obtained a phase diagram in the plane ``ion energy -- angle of
incidence'' it was shown that if the angle of incidence or ion
energy varies, then orientation of ripples can be changed. It is
important that in the considered interval of incidence angle, the obtained phase
diagram in figure~\ref{fig3} is topologically similar to the
experimental one. However, nonlinear KS equation~(\ref{Aeq1}) with
parameters defined by equations~(\ref{Aeq3})--(\ref{Aeq5}) does
not presume a stable smooth surface because $\nu_y$ is a negative
quantity.

\begin{figure}[ht]
\centering
\includegraphics[width=8cm]{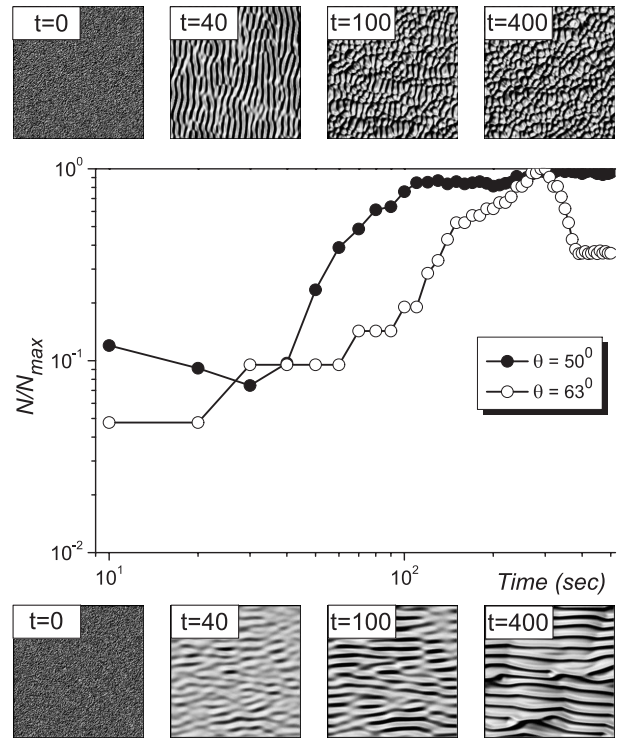}
\caption{Relative number of islands at $\ve=2$~keV and
characteristic snapshots at $\theta=50^{\circ}$ (top) and $\theta=63^{\circ}$
(bottom).} \label{fig4}
\end{figure}
To prove that holes and ripples are stable in time, let us consider the
dynamics of the surface morphology change. We analyze two
representative kinds of patterns shown in figure~\ref{fig3} as
snapshots A and E and compute the number of islands for each
pattern in time. To this end, we have cut the surface $h(x,y)$ at
an average height level $\langle h \rangle$ and calculated the
relative number of islands $N/N_{\rm max}$ at fixed times, where
$N_{\rm max}$ is a maximal value of islands. In our computation
scheme we used the following definition for the island: all points on
the surface with $h<\langle h \rangle$ belonging to one manifold
having a closed boundary, form an island. The corresponding boundary
of the island was obtained according to the percolation model
formalism. Results for relative number of islands were averaged
over 20 independent runs. Typical evolution of the number of islands is
shown in figure~\ref{fig4} at $\ve=2$~keV for $\theta=50^{\circ}$ and
$\theta=63^{\circ}$.
It is seen that the relative number of islands grows at small time
interval that corresponds to processes of the formation of islands. At
intermediate times, the relative number of islands decreases which
means a realization of coalescence processes. It is important that
in the process of ripple formation the coalescence regime is well
pronounced (see empty circles). On the contrary, for the process of
nanohole formation (filled circles), this such regime is only weakly
observed. At large times one has a stationary behavior of the
relative number of islands. Hence, processes of ripple and
nanohole formation are stationary ones: at large time intervals
the averaged number of islands does not change in time. Snapshots
of the silicon surface morphology for $\theta=50^{\circ}$ and
$\theta=63^{\circ}$ at $t=0$, $40$, $100$ and $400$ seconds are shown in
figure~\ref{fig4} in the top and in the bottom of the figure,
respectively.

\subsection{Wavelength dependence on the ion energy}

\begin{figure}[ht]
\centering
\includegraphics[width=8cm]{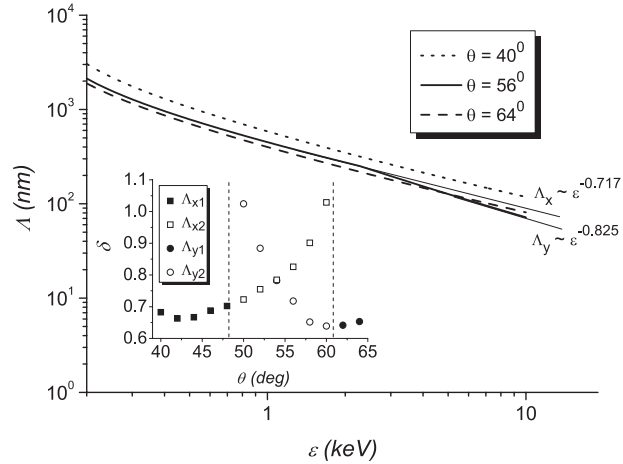}
\caption{Dependence of the wavelength on the ion energy (the
scaling exponent $\delta$ dependence on the incident angle is
shown in insertion).} \label{fig5}
\end{figure}

\looseness=-1Next, let us study the wavelength dependence on the incident ion
energy and on the angle of incidence. As it was shown earlier in the
Bradley-Harper theory, a relation between parameters $\nu_x$ and
$\nu_y$ determines the orientation of surface patterns. The
wavelength of selected patterns in the corresponding direction is
defined as follows: $\Lambda_{x,y}=2\pi \sqrt{2K/|\nu_{\min
x,y}|}$, where $\nu_{\min x,y} =\min(\nu_x\,,\nu_y)$. One needs to
note that following the phase diagram shown in figure~\ref{fig3}, a
variation in the ion energy at fixed angles of incidence causes a change in the orientation of structures: at small $\ve$ one has
structures elongated in $y$ direction, whereas at large $\ve$,
structures are horizontally elongated. Corresponding dependencies
of the wavelength versus ion energy at fixed values for incidence
angle are shown in figure~\ref{fig5}.
It is seen that the wavelength decreases with the ion energy
growth according to a power law and varies in the interval from
$100$~nm to $1~\mu$m. This result is in good correspondence with
experimental data for sputtering of the silicon target by Ar$^+$
ions~\cite{cite21}. It is principally important that as far as the
penetration depth depends on the ion energy in a nonlinear
manner (see the insert in figure~\ref{fig1}) one can expect a
deviation from the Bradley-Harper wavelength asymptote
$\Lambda\sim\ve^{-1/2}$. In figure~\ref{fig5} it is seen that at
small and large incidence angles (see dot and dash lines,
respectively) one has linear dependencies in log-log plot
characterized by the corresponding unique slope. However, at
intermediate values for $\theta$ related to the dash curve in
figure~\ref{fig3}, the dependence $\Lambda(\ve)$ has a kink. This
kink means a change in the orientation of patterns. In such a case
one has two slopes at small energies, i.e., before kink,  one has selected
the patterns characterized by $\Lambda_x$\,, whereas at large energies
the patterns are defined by $\Lambda_y$ (see solid line and
asymptotics in figure~\ref{fig5}). Therefore, at small ion energies
the patterns are oriented in $y$ direction, whereas at large ion
energies they are oriented in $x$ direction. Hence, for the
wavelength dependence on the ion energy one can write
$\Lambda\sim\ve^{-\delta}$ where the scaling exponent $\delta$ is
defined as a slope of the dependence $\Lambda(\ve)$ in double
logarithmic plot before and after the kink. The dependence of the
scaling exponent versus incidence angle is shown as an insert in
figure~\ref{fig5}. One can see that for the described interval for
the angle of incidence $\delta>1/2$. Moreover, at small and large
$\theta$ the exponent $\delta$ does not essentially change it values, whereas in the interval for $\theta$ when
$\Lambda(\ve)$ has a kink, the exponent $\delta$ varies from $0.65$
toward $1.05$. We should note that the obtained picture is realized
when the incoming ion flux $J$ and temperature $T$ are constants. In
the opposite case, variation in $J$ and $T$ leads to the known
asymptotes: $\Lambda\propto J^{-1/2}$, $\Lambda \propto T^{-1/2}
\exp (-E_a/2T)$, where $E_a$ is an activation energy.

\newpage
\subsection{Scaling properties of patterns}

Finally, let us study the scaling properties of the surface patterns,
computing growth and roughness exponents. To this end, we analyze a
height-height correlation function
$C_{h}(r,t)=\langle[h(r+r',t)-h(r',t)]^{2}\rangle$. In the
framework of dynamic scaling hypothesis following
references~\cite{cite23,cite24}, one arrives at scaling relations
$C_{h}(t)\propto t^{2\beta}$, $C_{h}(r)\propto r^{2\alpha}$,
allowing one to define the growth exponent $\beta$ and the
roughness exponent $\alpha$.

In reference~\cite{cite16} it was shown that there is a set of
exponents $\{\beta\}$ describing the universal behavior of the
correlation function at early stages of the system evolution. At
late times where a true scaling regime is observed there is a unique
value for $\beta$. The roughness exponent $\alpha$ takes similar
values at different time windows and can be considered as a
constant depending on the system parameters only. From practical
viewpoint, the analysis of the surface growth is urgent at large
time intervals where the true scaling regime is observed and there
is no essential difference in values $\beta$ at different time
windows. It is known that anisotropic surfaces studied in this
paper may exhibit a more complex dynamic scaling behaviour than
isotropic ones because anisotropy of the surface is reflected
in lateral correlations of the surface
roughness~\cite{cite36,cite37}. In reference~\cite{cite26} it was
proposed to use local roughness scales $\alpha_x$\,, $\alpha_y$ in
the directions normal and parallel to the projection direction of
the ion beam. Values for growth and roughness exponents together
with surface tensions $\nu_x$ and $\nu_y$ at $\ve=2$~keV and fixed
values for incidence angle are presented in table~\ref{table1}.
It is seen that when $\nu_x<\nu_y$, a relation $\alpha_x>\alpha_y$
is realized due to orientation of the structures in $y$
direction. On the contrary, if $\nu_x>\nu_y$ holds, then one has
$\alpha_x<\alpha_y$\,. Hence, making an analysis of the obtained
scaling exponents, one can conclude that if structures are
oriented in $y$ direction, then roughness is larger in
$x$-direction and \emph{vice versa} (compare patterns in snapshots
A, C, D and E in figure~\ref{fig3} with exponents in
table~\ref{table1}). The obtained results for growth and roughness
exponents are in good correspondence with experimental studies of
the silicon target sputtered by Ar$^+$ ions
(see~\cite{cite26,cite35}).
\begin{table}[ht]
\caption{Growth and roughness scaling exponents at $\ve=2$~keV.}
\label{table1} 
\begin{center}
\renewcommand{\arraystretch}{0}
\begin{tabular}{p{0.2in}p{0.6in}p{0.6in}p{0.6in}p{0.5in}p{0.5in}p{0.5in}} \hline \hline
 &$\theta$ & $\alpha_x$ & $\alpha_y$ & $\beta$ & $\nu_x$ & $\nu_y$\\
 \hline \hline
 &$50^{\circ}$ & 0.90 & 0.82 & 0.23 & --0.222 & --0.151\\
 &$55^{\circ}$ & 0.94 & 0.90 & 0.22 & --0.137 & --0.127\\
 &$58^{\circ}$ & 0.90 & 0.95 & 0.21 & --0.067 & --0.112\\
 &$63^{\circ}$ & 0.89 & 0.99 & 0.17 & \ 0.086  & --0.087\\
 \hline \hline
\end{tabular}
\renewcommand{\arraystretch}{1}
\end{center}
\end{table}

\section{Conclusions}

Two-level modeling for nanoscale pattern formation on silicon
target induced by Ar$^+$ ion sputtering has been reported. We have used Monte-Carlo simulations and a continuum approach based on the
Bradley-Harper theory. It was shown that for the described system, the
dependencies of the averaged penetration depth of the incident ion
and the corresponding distribution widths of the deposited energy in
directions parallel and perpendicular to the incoming beam versus
ion energy are of the power-law form. Varying the incoming ion energy
and ion incidence angle, we have defined the sputtering yield with the
help of Monte-Carlo simulations. The obtained results have been used
in the modified Bradley-Harper theory within the framework of
two-scale modeling scheme.

We have computed a phase diagram for control parameters: i.e., incidence
angle and ion energy that defines possible patterns on silicon
target sputtered by Ar$^+$ ions. It was shown that at small
incidence angles, nanohole patterns are realized, whereas at large
incidence angles, pattern type of ripples is observed. Analyzing
the morphology change of silicon surface we have shown that during
the system evolution, the number of nanoholes/ripples becomes constant, indicating stability of the obtained structures in time.

We have found that there are deviations from the Bradley-Harper asymptotics for
the wavelength dependence on the ion energy. Moreover, when the orientation of
patterns changes, a kink is realized in such asymptotics. The exponent of
such power-law asymptotics depends on the angle of incidence. At fixed values
for incidence angle one has two scaling exponents related to small and large
values for the ion energy according to a change in the orientation of structures.
While studying the scaling characteristics of the height-height correlation function, the
growth exponent together with longitudinal and transverse roughness exponents
are obtained for different values of incidence angle at a fixed ion energy. It
was shown that relations between roughness exponents are defined through
relations between corresponding effective surface tensions.

The results obtained in a two-scale modeling scheme are in good
correspondence with the known theoretical and experimental data for
sputtering of silicon target by Ar$^+$ ions in the considered
interval of values for incidence angle of ions, the incoming ion energy,
temperature and ion flux~\cite{cite21,cite22,cite26,cite6,cite35}.

\newpage
\ukrainianpart

\title{Зміна морфології поверхні кремнію при розпиленні його іонами аргону}%
\author{В.О. Харченко, Д.О. Харченко}
\address{Інститут прикладної фізики НАН України, вул. Петропавлівська 58, 40030 Суми, Україна}
\makeukrtitle
\begin{abstract}
\tolerance=3000%
Проводиться теоретичне дослідження процесів зміни морфології поверхні кремнію
при розпиленні його іонами аргону в рамках дворівневої схеми, що враховує
методи Монте-Карло та модифіковану теорію Бредлі-Харпера. Отримано та
проаналізовано фазову діаграму у площині кут падіння налітаючого іону та
енергія іону, що ілюструє можливі типи поверхневих нано-структур. Отримано
узагальнену степеневу залежність довжини хвилі отриманих поверхневих структур
від енергії налітаючих іонів. Проаналізовано показник росту і повздовжній та
поперечний показники шорсткості отриманих поверхонь.
\keywords іонне розпилення, морфологія поверхні, нано-структури

\end{abstract}

\label{last@page}
\end{document}